\documentclass[12pt]{article}
\usepackage{graphicx}
\begin {document}
\title {The birth of a ghost star}

\author{Luis Herrera \\Instituto Universitario de F\'isica
Fundamental y Matem\'aticas, \\Universidad de Salamanca, Salamanca 37007, \\Spain \thanks{ E-mail address: lherrera@usal.es}\\Alicia Di Prisco \\Escuela de F\'{\i}sica. Facultad de Ciencias.\\ Universidad Central de Venezuela. Caracas, Venezuela. \\Justo Ospino \\Departamento de Matem\'atica Aplicada\\ Instituto Universitario de F\'isica
Fundamental y Matem\'aticas, \\Universidad de Salamanca, \\Salamanca 37007, Spain}

\date{}
\maketitle

\begin{abstract}
 We present  a model of an evolving spherically symmetric dissipative self-gravitating fluid distribution which tends asymptotically to a ghost star, meaning  that the end state of such a system corresponds to a static fluid distribution with vanishing total mass, and energy-density distribution which is negative in some regions of the fluid. The model is inspired in a solution representing a fluid evolving quasi-homologously  and with vanishing complexity factor. However in order to satisfy the asymptotic behavior mentioned above, the starting solution has to be  modified, as a consequence of which   the resulting model only satisfies the two previously mentioned conditions, asymptotically. Additionally a condition on the variation of the infinitesimal proper radial
distance between two neighboring points  per unit of proper time is imposed, which implies the presence of a cavity surrounding the center. Putting together all these conditions we are able to obtain an analytical model  depicting the emergence of a ghost star. Some potential observational consequences of this phenomenon are briefly discussed at the last section.
\end{abstract}

\newpage

\section{Introduction}

In a recent paper \cite{1n} the   concept of ghost star was introduced and studied in detail. Such a concept, inspired  in early ideas by Zeldovich \cite{2n,2nb}, describes fluid distributions which do not produce gravitational field outside its boundary surface (i.e. its total mass vanishes). In order to achieve the vanishing of the total mass (for a non-trivial fluid distribution) one must assume the existence of some regions within the fluid sphere endowed with negative energy-density. Some examples of this kind of fluid distribution may be found in \cite{1n,3n,4n} (see also \cite{na} for more recent developments).

 More recently we have studied solutions  which, either correspond to the  adiabatic evolution  of  a ghost star, or describe the evolution of fluid distributions
which attain the ghost star status momentarily at some point of their lives, abandoning such state immediately  afterward \cite{5n}. 

It should be stressed that the term ``Ghost stars'' , comes from the analogy with   some Einstein--Dirac  neutrinos (named  ghost neutrinos)  which do not produce gravitational field but  still are characterized by  non vanishing current density \cite{gn,gn1,gn2}. Thus, any confusion with the same terminology used in quantum field theory should be dismissed.

However neither of the models exhibited in the references above describe solutions  leading asymptotically (as $t\rightarrow \infty$) to a ghost star.

It is the purpose of this work to present a model of an evolving fluid distribution describing the emergence of  a ghost star, as the end point of its evolution.

In order to obtain our model we have initially started by imposing three conditions
\begin{itemize} 
\item Vanishing of the complexity factor \cite{6n,7n}
\item Quasi-homologous evolution (QH)\cite{7nn}
\item The variation of the infinitesimal proper radial
distance between two neighboring points per unit of
proper time vanishes \cite{9n}.
\end{itemize}
The first two conditions have been shown to be useful in the description of  the structure and evolution of self--gravitating fluids. The second one represents a generalization of the well known homologous evolution in Newtonian hydrodynamics \cite{10n,11n,12n}. 

The third condition implies the existence of a cavity surrounding the center, and therefore appears to be a useful tool for the modeling of cosmic voids \cite{18n,19n}.

Notwithstanding, we have resorted to the above conditions for purely heuristic reasons, its physical interest being, in the context of this work,  a fact of  secondary relevance. 

The solution obtained under the three conditions above (hereafter referred to as ``primeval solution'') does  not satisfy the asymptotic conditions required to obtain a static ghost star as the end point of the evolution. Accordingly we have modified this solution in order to satisfy the conditions ensuring the formation of a ghost star.

The final solution matches smoothly on the external boundary surface  with the Minkowski space-time as $t\rightarrow \infty$. Instead, matching conditions are not satisfied on the boundary surface delimiting the fluid distribution from the inside (not even asymptotically), accordingly we have a thin shell on that surface. 

The physical properties of the model will be analyzed in detail and the characteristics of the ghost star appearing at the end of the evolution  will be discussed.

\section{THE GENERAL SETUP OF THE PROBLEM: NOTATION, VARIABLES AND EQUATIONS}

We consider a spherically symmetric distribution  of  fluid, which is  bounded from the outside by a spherical surface $\Sigma^{(e)}$ and, since we shall assume a cavity  surrounding the center, the fluid  is also bounded from inside by a spherical surface   $\Sigma^{(i)}$. The matter content consists in a locally anisotropic fluid (principal stresses unequal)  undergoing dissipation in the
form of heat flow (diffusion approximation). 

Thus, in comoving coordinates,  the general line element may be written as
\begin{equation}
ds^2=-A^2dt^2+B^2dr^2+R^2(d\theta^2+\sin^2\theta d\phi^2),
\label{1}
\end{equation}
where the functions $A, B, R$ depend on $t$ and $r$.

The energy--momentum tensor  takes the form
\begin{eqnarray}
T_{\alpha\beta}&=&(\mu +
P_{\perp})V_{\alpha}V_{\beta}+P_{\perp}g_{\alpha\beta}+(P_r-P_{\perp})\chi_{
\alpha}\chi_{\beta}\nonumber \\&+&q_{\alpha}V_{\beta}+V_{\alpha}q_{\beta}
, \label{3}
\end{eqnarray}
where $\mu$ is the energy density, $P_r$ the radial pressure,
$P_{\perp}$ the tangential pressure, $q^{\alpha}$ the heat flux, $V^{\alpha}$ the four velocity of the fluid,
and $\chi^{\alpha}$ a unit four vector along the radial direction. These quantities
satisfy
\begin{eqnarray}
V^{\alpha}V_{\alpha}=-1, \;\; V^{\alpha}q_{\alpha}=0, \;\; \chi^{\alpha}\chi_{\alpha}=1,\;\;
\chi^{\alpha}V_{\alpha}=0.
\end{eqnarray}

It will be convenient to express the  energy momentum tensor  (\ref{3})  in the equivalent (canonical) form
\begin{equation}
T_{\alpha \beta} = {\mu} V_\alpha V_\beta + P h_{\alpha \beta} + \Pi_{\alpha \beta} +
q \left(V_\alpha \chi_\beta + \chi_\alpha V_\beta\right), \label{Tab}
\end{equation}
with
$$ P=\frac{P_{r}+2P_{\bot}}{3}, \qquad h_{\alpha \beta}=g_{\alpha \beta}+V_\alpha V_\beta,$$

$$\Pi_{\alpha \beta}=\Pi\left(\chi_\alpha \chi_\beta - \frac{1}{3} h_{\alpha \beta}\right), \qquad \Pi=P_{r}-P_{\bot}.$$

Since we are considering comoving observers, we have
\begin{eqnarray}
V^{\alpha}&=&A^{-1}\delta_0^{\alpha}, \;\;
q^{\alpha}=qB^{-1}\delta^{\alpha}_1, \;\;
\chi^{\alpha}=B^{-1}\delta^{\alpha}_1. \label{k}
\end{eqnarray}

It is worth noticing that we do not add explicitly  bulk or shear viscosity to the system because they
can be trivially absorbed into the radial and tangential pressures, $P_r$ and
$P_{\perp}$, of the collapsing fluid (in $\Pi$). Also we do not explicitly  introduce  dissipation in the free streaming approximation since it can be absorbed in $\mu, P_r$ and $q$. 

\subsection{Einstein equations}
\label{sec:2}
Einstein's field equations for the interior spacetime (\ref{1}) are given by
\begin{equation}
G_{\alpha\beta}=8\pi T_{\alpha\beta}.
\label{2}
\end{equation}
The non null components of (\ref{2})
with (\ref{1}) and  (\ref{3}),
read

\begin{eqnarray}
8\pi T_{00}=8\pi  \mu A^2
=\left(2\frac{\dot{B}}{B}+\frac{\dot{R}}{R}\right)\frac{\dot{R}}{R}
-\left(\frac{A}{B}\right)^2\left[2\frac{R^{\prime\prime}}{R}+\left(\frac{R^{\prime}}{R}\right)^2
-2\frac{B^{\prime}}{B}\frac{R^{\prime}}{R}-\left(\frac{B}{R}\right)^2\right],
\label{T00} 
\end{eqnarray}

\begin{eqnarray}
8\pi T_{01}=-8\pi  q AB=-2\left(\frac{{\dot
R}^{\prime}}{R}-\frac{\dot B}{B}\frac{R^{\prime}}{R}-\frac{\dot
R}{R}\frac{A^{\prime}}{A}\right),
\label{T01} 
\end{eqnarray}

\begin{eqnarray}
8\pi T_{11}=
 8\pi P_r B^2
=-\left(\frac{B}{A}\right)^2\left[2\frac{\ddot{R}}{R}-\left(2\frac{\dot
A}{A}-\frac{\dot{R}}{R}\right)
\frac{\dot R}{R}\right]
+\left(2\frac{A^{\prime}}{A}+\frac{R^{\prime}}{R}\right)\frac{R^{\prime}}{R}-\left(\frac{B}{R}\right)^2,
\label{T11} 
\end{eqnarray}

\begin{eqnarray}
8\pi T_{22}&=&\frac{8\pi}{\sin^2\theta} T_{33}
=8\pi P_{\perp}R^2
=-\left(\frac{R}{A}\right)^2\left[\frac{\ddot{B}}{B}+\frac{\ddot{R}}{R}
-\frac{\dot{A}}{A}\left(\frac{\dot{B}}{B}+\frac{\dot{R}}{R}\right)
+\frac{\dot{B}}{B}\frac{\dot{R}}{R}\right]
\nonumber \\&+&\left(\frac{R}{B}\right)^2\left[\frac{A^{\prime\prime}}{A}
+\frac{R^{\prime\prime}}{R}-\frac{A^{\prime}}{A}\frac{B^{\prime}}{B}
+\left(\frac{A^{\prime}}{A}-\frac{B^{\prime}}{B}\right)\frac{R^{\prime}}{R}\right],\label{T22}
\end{eqnarray}

where dots and primes denote derivative with respect to $t$ and $r$ respectively.

\subsection{Kinematical variables and the mass function}
\label{sec:3}

The three non-vanishing kinematical variables are the four--acceleration  $a_{\alpha}$, the expansion scalar $\Theta$ and the shear tensor $\sigma_{\alpha \beta}$. 
The corresponding expressions follow at once from their definitions. 

Thus
\begin{equation}
a_{\alpha}=V_{\alpha ;\beta}V^{\beta}, \label{4b}
\end{equation}
producing 

\begin{equation}
a_1=\frac{A^{\prime}}{A}, \;\;
a^2=a^{\alpha}a_{\alpha}=\left(\frac{A^{\prime}}{AB}\right)^2,
\label{5c}
\end{equation}
with  $a^\alpha= a \chi^\alpha$.

The expansion $\Theta$ is given by 

\begin{equation}
\Theta={V^{\alpha}}_{;\alpha}=\frac{1}{A}\left(\frac{\dot{B}}{B}+2\frac{\dot{R}}{R}\right),\label{th}
\end{equation}
and for the shear  tensor we have

\begin{equation}
\sigma_{\alpha\beta}=V_{(\alpha
;\beta)}+a_{(\alpha}V_{\beta)}-\frac{1}{3}\Theta h_{\alpha\beta}, \label{4a}
\end{equation}
with only one  non--vanishing independent component.

Using  (\ref{k}) and  (\ref{4a}) we may write
\begin{equation}
\sigma_{\alpha \beta}=\sigma\left(\chi_\alpha \chi_\beta-\frac{h_{\alpha \beta}}{3}\right),
\label{sigma}
\end{equation}
where
\begin{equation}
\sigma=\frac{1}{A}\left(\frac{\dot{B}}{B}-\frac{\dot{R}}{R}\right).\label{5b1}
\end{equation}
\subsection{The mass function}
\label{sec:4}
Next, the mass function $m(t,r)$ introduced by Misner and Sharp \cite{13n,13nn}
is given by 

\begin{equation}
m(t,r)=\frac{R^3}{2}{R_{23}}^{23} 
=\frac{R}{2}\left[\left(\frac{\dot{R}}{A}\right)^2
-\left(\frac{R^{\prime}}{B}\right)^2+1\right].
 \label{18}
\end{equation}

To study the dynamical properties of the system, let us  introduce,
following Misner and Sharp  the proper time derivative $D_T$
given by
\begin{equation}
D_T=\frac{1}{A}\frac{\partial}{\partial t}, \label{16}
\end{equation}
and the proper radial derivative $D_R$,
\begin{equation}
D_R=\frac{1}{R^{\prime}}\frac{\partial}{\partial r}.\label{23a}
\end{equation}

Using (\ref{16}) we can define the velocity $U$ of the collapsing
fluid as the variation of the ``areal''  radius ($R$) with respect to proper time, i.e.
\begin{equation}
U=D_TR.  \label{19}
\end{equation}
Then (\ref{18}) can be rewritten as
\begin{equation}
E \equiv \frac{R^{\prime}}{B}=\left[1+U^2-\frac{2m(t,r)}{R} \right]^{1/2}.
\label{20}
\end{equation}

From (\ref{18}) we may easily obtain
\begin{eqnarray}
D_Rm=4\pi\left( \mu+  q \frac{U}{E}\right)R^2 .
\label{27b}
\end{eqnarray}
Equation (\ref{27b}) may be integrated to obtain
\begin{equation}
m=\int^{r}_{0}4\pi R^2 \left( \mu +  q \frac{U}{E}\right)R^\prime dr, \label{27int}
\end{equation}
(assuming a regular centre to the distribution, so $m(0)=0$), 
or
\begin{equation}
\frac{3m}{R^3} =4 \pi  \mu - \frac{4 \pi}{R^3}\int^r_0{R^3  \mu^\prime  dr} + \frac{4 \pi}{R^3}\int^r_0{3  q \frac{U}{E}R^2 R^\prime dr} \label{3mi}.
\end{equation}

\subsection{The junction conditions}
\label{sec:2}
Outside $\Sigma^{(e)}$  we have the Vaidya
spacetime (or Schwarzschild in the dissipationless case),
described by
\begin{equation}
ds^2=-\left[1-\frac{2M(v)}{r}\right]dv^2-2drdv+r^2(d\theta^2
+\sin^2\theta
d\phi^2) \label{19d},
\end{equation}
where $M(v)$  denotes the total mass,
and  $v$ is the retarded time.
The matching of the non-adiabatic sphere to
the Vaidya spacetime, on the surface $r=r_{\Sigma^{(e)}}=$ constant, in the absence of thin shells (Darmois conditions \cite{14nn}, see also \cite{chan1}), implies   the continuity of the first and the second fundamental forms through the matching hypersurface, producing
\begin{equation}
m(t,r)\stackrel{\Sigma^{(e)}}{=}M(v), \label{20}
\end{equation}
and
\begin{equation}
q\stackrel{\Sigma^{(e)}}{=}P_r  \label{20lum}.
\end{equation}

In the case when a cavity forms we also have to match the solution to the Minkowski spacetime on the boundary surface delimiting the empty cavity ($\Sigma^{(i)}$). In this case the matching conditions imply
\begin{equation}
m(t,r)\stackrel{\Sigma^{(i)}}{=}0, \label{junction1i}
\end{equation}

\begin{equation}
q\stackrel{\Sigma^{(i)}}{=}P_r\stackrel{\Sigma^{(i)}}{=}0.\label{j3in}
\end{equation}

As we shall see below, in our  model the Darmois conditions  cannot be satisfied  on $\Sigma^{(i)}$, in which case we must  allow  the presence of thin shells on $\Sigma^{(i)}$, implying   discontinuities in the mass function \cite{15nn}. 

On the other hand Darmois conditions are satisfied on $\Sigma^{(e)}$ but only asymptotically (as $t\rightarrow\infty$). In other words a thin shell is present on $\Sigma^{(e)}$  during the evolution, disappearing as the ghost star forms.

\subsection{THE TRANSPORT EQUATION}
\label{sec:4}
 In the diffusion approximation we shall need a transport equation to evaluate the temperature and its evolution within the fluid distribution. Here we shall resort to a transport equation derived from a causal  dissipative theory ( e.g. the
Israel-Stewart second
order phenomenological theory for dissipative fluids \cite{14n,15n,16n}).

Thus the  corresponding  transport equation for the heat flux reads
\begin{equation}
\tau
h^{\alpha\beta}V^{\gamma}q_{\beta;\gamma}+q^{\alpha}=-\kappa h^{\alpha\beta}
(T_{,\beta}+Ta_{\beta}) -\frac 12\kappa T^2\left( \frac{\tau
V^\beta }{\kappa T^2}\right) _{;\beta }q^\alpha ,  \label{21t}
\end{equation}
where $\kappa $  denotes the thermal conductivity, and  $T$ and
$\tau$ denote temperature and relaxation time respectively. Observe
that, due to the symmetry of the problem, equation (\ref{21t}) only
has one independent component, which may be written as
\begin{equation}
\tau{\dot q}=-\frac{1}{2}\kappa qT^2\left(\frac{\tau}{\kappa
T^2}\right)^{\dot{}}-\frac{1}{2}\tau q\Theta A-\frac{\kappa}{B}(TA)^{\prime}-qA.
\label{te}
\end{equation}
In the case $\tau=0$ we recover the Eckart--Landau equation \cite{17,67}, and in the Newtonian limit we recover the Cattaneo equation \cite{co,Ref9,Ref10}

 For simplicity we shall consider here the so called ``truncated'' version where the last term in (\ref{21t}) is neglected \cite{17n},
\begin{equation}
\tau
h^{\alpha\beta}V^{\gamma}q_{\beta;\gamma}+q^{\alpha}=-\kappa h^{\alpha\beta}
(T_{,\beta}+Ta_{\beta}) \label{V1},
\end{equation}
and whose    only non--vanishing independent component becomes
\begin{equation}
\tau \dot q+qA=-\frac{\kappa}{B}(TA)^{\prime}. \label{V2}
\end{equation}

\section{Three conditions leading to our model}
As mentioned before we shall start the building up of our model by imposing three conditions on the fluid distribution, these are: the vanishing complexity factor condition, the quasi--homologous condition and a kinematical condition  on the variation of the infinitesimal proper radial
distance between two neighboring points  per unit of proper time. In what follows we shall briefly describe these conditions.

\subsection{The vanishing   complexity factor condition}
The complexity factor is  a scalar function that has been proposed in order  to measure the degree of complexity of a given fluid distribution \cite{6n, 7n}.

The complexity factor is identified with the scalar function $Y_{TF}$ which defines the trace--free part of the electric Riemann tensor (see \cite{7n} for details).

It can be expressed in terms of physical variables as 

\begin{equation}
Y_{TF}= - 8 \pi \Pi   + \frac{4 \pi}{R^3}\int^r_0{R^3\left(\mu^\prime - \frac{3  q BU}{R}\right) dr},
\label{Yi}
\end{equation}
or in  terms of the metric functions

\begin{eqnarray}
Y_{TF}= \frac{1}{A^2}\left[\frac{\ddot R}{R} - \frac{\ddot B}{B} + \frac{\dot A}{A}\left(\frac{\dot B}{B} - \frac{\dot R}{R}\right)\right]+ \frac{1}{ B^2} \left[\frac{A^{\prime\prime}}{A} -\frac{A^{\prime}}{A}\left(\frac{B^{\prime}}{B}+\frac{R^{\prime}}{R}\right)\right] .
\label{itfm}
\end{eqnarray}

We shall impose the vanishing of the complexity factor in order to find an analytical solution, however as we shall see below, such a solution does not satisfy the required asymptotic behavior. In order to obtain a model with the appropriate asymptotic behavior we shall modify this primeval solution, as a consequence of which the resulting model will satisfy the vanishing complexity factor condition only asymptotically. 

\subsection{The quasi-homologous condition}
\label{sec:3}

The QH condition is a generalization of the homologous condition (H) which has been assumed in \cite{7n} to represent  the simplest mode of evolution of the fluid distribution. However this last condition appears  to be too   stringent thereby  excluding many potential interesting scenarios. Therefore in \cite{7nn} we have proposed to relax (H), and replaced it by what we  called  the  ``quasi--homologous'' condition (QH). 

More specifically, the H condition implies that 
\begin{equation}
U=\tilde a(t)R, \qquad \tilde a(t)\equiv\frac{U_{\Sigma^{(e)}}}{R_\Sigma^{(e)}},
\label{h1n}
\end{equation}
and 
 
 \begin{equation}
\frac{R_I}{R_{II}}=\mbox{constant},
\label{hn2}
\end{equation}
where $R_I$ and $R_{II}$ denote the areal radii of two concentric shells ($I,II$) described by $r=r_I={\rm constant}$, and $r=r_{II}={\rm constant}$, respectively. 
 
 These  relationships  are characteristic of the homologous evolution in Newtonian hydrodynamics \cite{10n,11n,12n}. More so, in this latter case (\ref{h1n}) implies (\ref{hn2}). However in the relativistic case both (\ref{h1n}) and  (\ref{hn2}) are in general independent, and the former implies the latter only in very special cases.
 
On the other hand QH only requires (\ref{h1n}), which using the field equations may  also be written as (see \cite{7nn} for details)
\begin{equation}
\frac{4\pi}{R^\prime}B  q+\frac{\sigma}{ R}=0.
\label{ch1}
\end{equation}

As  already mentioned we shall start the building up of our model by assuming that the evolution of the fluid distribution proceeds in the quasi-homologous regime (QH). Since such a condition leads to an asymptotic behavior which is incompatible with the formation of a ghost star, we should modify the primeval solution. As a consequence of this modification the final solution will not satisfy the (QH) except in the static limit $t\rightarrow \infty$, when it is trivially satisfied.

\subsection{A kinematical restriction}
\label{sec:6}
In order to obtain our primeval model,  besides the condition of the vanishing complexity factor and the quasi--homologous evolution, we shall impose a condition on a kinematical variable.  For doing that let us first   introduce another concept of velocity, different from $U$, which measures  the variation of the infinitesimal proper radial distance between two neighboring points ($\delta l$) per unit of proper time, i.e. $D_T(\delta l)$.
Thus, it can be shown that (see \cite{9n,18} for details)
\begin{equation}
 \frac{D_T(\delta l)}{\delta l}= \frac{1}{3}(2\sigma +\Theta),
\label{vel15}
\end{equation}
or,
\begin{equation}
 \frac{D_T(\delta l)}{\delta l}= \frac{\dot B}{AB}.
\label{vel}
\end{equation}

As an additional restriction we shall assume  $D_T(\delta l)=0$, in which case   $B=B(r)$, from which  a reparametrization of the coordinate $r$ allows us to write without loss of generality $B=1$ implying $R^\prime=E$, and as it follows from (\ref{th}) and (\ref{5b1})
\begin{equation}
\sigma=-\frac{U}{R}=-\frac{\Theta}{2}.
\label{nk1}
\end{equation}

Since the center of symmetry ($r=0$) does not move all along the evolution, it appears evident that any evolving fluid satisfying the condition $B=1$ cannot fill the central region. Therefore we shall assume the center to be surrounded by a void cavity  with  boundary surface $\Sigma^{(i)}$, whose areal radius changes in such a way that $D_T(\delta l)=0$ for all fluid elements. 

From the comment above  it should be clear why this condition has been considered in the past as a useful tool for describing galactic voids. However we should stress the fact that here we are adopting this kinematical condition   just as a heuristic  hypothesis, in order to obtain an analytical model describing the emergence of a ghost star.

In the next section we shall build up  a model leading to a ghost star.

\section{Building up The model}
\label{sec:7}

We shall now proceed to find a model giving rise to a ghost star. This will be achieved in three steps. First we shall find a primeval solution satisfying the three conditions:  $Y_{TF}=0$, (\ref{ch1}) and $B=1$. In a second step we shall modify this primeval solution in order to satisfy  the desired asymptotic behavior. Finally, in the third step we shall make a specific choice of some arbitrary functions and constants to fully determine the model. This final model satisfies the condition $B=1$, but conditions  $Y_{TF}=0$ and (\ref{ch1}) are satisfied only asymptotically.

\subsection{The primeval solution}
\label{sec:9}
\noindent Let us start by considering a model satisfying the constraint  $B=1$. This model is endowed with a cavity surrounding the center, accordingly we should not worry about regularity  conditions at the center.

In this case the physical variables read

\begin{equation}
  8\pi \mu = \frac{1}{A^2}\frac{\dot{R}^2}{R^2}-\frac{2R^{\prime\prime}}{R}-\frac{R^{\prime 2}}{R^2}+\frac{1}{R^2}, \label{mu}
  \end{equation}

  \begin{equation}\label{Pr}
    8\pi P_r = -\frac{1}{A^2}\left( \frac{2\ddot{R}}{R}-\frac{2\dot{A}}{A}\frac{\dot{R}}{R}+\frac{\dot{R}^2}{R^2}\right)
  +\frac{2A^\prime}{A}\frac{R^\prime}{R}+\frac{{R^{\prime}}^2}{R^2}-\frac{1}{R^2},
  \end{equation}

  \begin{equation}\label{Pt}
   8\pi P_\perp =  -\frac{1}{A^2}\left ( \frac{\ddot{R}}{R}-\frac{\dot{A}}{A}\frac{\dot{R}}{R}\right )+\frac{A^{\prime\prime}}{A}+\frac{R^{\prime\prime}}{R}+\frac{A^\prime}{A}\frac{R^\prime}{R},
  \end{equation}
  \begin{equation}\label{fq}
  4\pi q =\frac{1}{A}\left (\frac{\dot{R}^\prime}{R}-\frac{A^\prime}{A}\frac{\dot{R}}{R}\right ) = -\sigma^{\prime}-\sigma \frac{R^\prime}{R},
\end{equation}
and for the kinematical variables we have

\begin{equation}\label{VaCi}
\sigma =-\frac{\dot{R}}{AR},\qquad \Theta=\frac{2\dot{R}}{AR}.
\end{equation}

Next, imposing the quasi--homologous condition, we obtain

\begin{equation}\label{CoHo}
  U=\tilde{a}(t) R\quad \Rightarrow\qquad \tilde{a}(t)=\frac{\dot{R}}{AR}\quad\Rightarrow \qquad \sigma=-\tilde{a}(t).
\end{equation}
In other words, the QH condition implies that in this case $\sigma$ only depends on $t$.

On the other hand the condition $Y_{TF}=0$ produces 
\begin{eqnarray}
  Y_{TF} &=& \frac{1}{A^2} \left(\frac{\ddot{R}}{R}-\frac{\dot{A}}{A}\frac{\dot{R}}{R}\right)+\frac{A^{\prime \prime}}{A}-\frac{A^\prime}{A}\frac{R^\prime}{R}\\
   &=&\sigma^2-\frac{\dot{\sigma}}{A} +\frac{A^{\prime \prime}}{A}-\frac{A^\prime}{A}\frac{R^\prime}{R}=0.
\end{eqnarray}

\noindent Thus, the conditions of vanishing complexity factor, $B=1$ and quasi--homologous evolution read
\begin{equation}\label{Cond1}
  A^{\prime\prime}-\frac{A^\prime R^\prime}{R}+A\sigma^2=\dot{\sigma},
\end{equation}
and
\begin{equation}\label{Cond2}
  \frac{\dot{R}}{R}=-\sigma A,
\end{equation}
respectively, with $\sigma=\sigma(t)$.

In order to solve the above system of equations it would be useful to introduce the intermediate variables $(X, Y)$,
\begin{equation}\label{ccam}
  A=X+\frac{\dot{\sigma}}{\sigma^2}\quad  {\rm and} \quad R=X^\prime Y,
\end{equation}

\noindent in terms of which  (\ref{Cond1}) and  (\ref{Cond2}), become

\begin{equation}\label{Cond1P}
  -\frac{X^\prime}{X}\frac{Y^\prime}{Y}+\sigma^2=0,
\end{equation}

\begin{equation}\label{Cond2P}
  \frac{\dot X^\prime}{X^\prime}+\frac{\dot Y}{Y}=-\sigma X-\frac{\dot{\sigma}}{\sigma}.
\end{equation}

In what follows we shall impose an additional restriction to solve the above system, specifically we shall assume that $X$ is a separable function, i.e. 

\begin{equation}\label{VSX}
  X=\tilde X(r) {\cal T}(t).
\end{equation}
\noindent Then  feeding back (\ref{VSX}) into (\ref{Cond1P}), and taking $t$-derivative we obtain
\begin{equation}\label{Cond1PV}
  -\frac{\tilde{X^\prime}}{\tilde{X}}\left (\frac{\dot Y}{Y}\right )^\prime+2\sigma\dot{\sigma}=0.
\end{equation}

\noindent Likewise, feeding back (\ref{VSX}) into  (\ref{Cond2P}) and taking the $r$-derivative we obtain
\begin{equation}\label{Cond2PV}
  \left (\frac{\dot Y}{Y}\right )^\prime=-\sigma \tilde{X}^\prime {\cal T}.
\end{equation}

\noindent The combination of (\ref{Cond1PV}) and (\ref{Cond2PV}) produces
\begin{equation}\label{Cond12PV}
  \frac{\tilde X^{\prime 2}}{\tilde X}=-\frac{2\dot{\sigma}}{{\cal T}}\equiv \beta^2,
\end{equation}
\noindent  where $\beta$ is a constant. 

Then, from the integration of (\ref{Cond12PV}) we have
\begin{equation}\label{solPV}
  \tilde{X}=\frac{(\beta r+c_1)^2}{4}\quad {\rm and}  \quad {\cal T}(t)=-\frac{2\dot{\sigma}}{\beta ^2},
\end{equation}
where $c_1$ is a constant of integration.

\noindent  Thus, the metric functions  become
\begin{eqnarray}
  A &=& \frac{\dot{\sigma}}{2\beta^2\sigma^2} \left [2\beta^2-\sigma^2 (\beta r+c_1)^2\right],\label{fume1n}\\
  R &=& F(t) g(r)(\beta r+c_1)e^{\frac{\sigma ^2 }{4\beta^2}(\beta r+c_1)^2},\label{fumeI}
\end{eqnarray}
\noindent where $F(t)$ and $g(r)$ are two  arbitrary functions of their arguments. This a generalized version of the solution exhibited in Sec.7.2.1 in \cite{7nn}.

However, functions $F(t)$ and $g(r)$ are not completely arbitrary. Indeed,  taking the $t$ derivative of (\ref{fumeI}) and feeding it back into (\ref{Cond2}) we obtain 
\begin{equation}
\frac{\dot F}{F}=-\frac{\dot \sigma}{\sigma},
\label{exp1}
\end{equation}
producing
\begin{equation}
F=\frac{c_2}{\sigma},
\label{exp2}
\end{equation}
where $c_2$ is an arbitrary constant.

The above result implies that in the static limit (when  $\sigma=0$), $F\rightarrow \infty$, leading to $R\rightarrow \infty$  in that limit.
 
 Still worse from the above and (\ref{fumeI}) it follows that 
 \begin{equation}
\dot R=-\frac{c_2 \dot\sigma g(\beta r+c_1)}{\sigma^2}.
\label{exp3}
\end{equation}

Thus if we want our system to  be static in the limit $t\rightarrow \infty$, we should demand $\frac{ \dot\sigma}{\sigma^2}\rightarrow 0$ as $t\rightarrow \infty$. But such a condition would imply because of (\ref{fume1n}) that $A\rightarrow 0$ as $t\rightarrow \infty$, which of course in unacceptable.

On the other hand using (\ref{fume1n}) and  (\ref{fumeI}), the vanishing complexity factor condition (\ref{Cond1}) reads
 \begin{equation}
\frac{\dot\sigma g^\prime (\beta r+c_1)}{\beta g}=0,
\label{exp4}
\end{equation}
implying $g=constant$. However, as we shall see below, we shall need $g=g(r)$ in order to satisfy matching conditions. 

In other words the metric functions (\ref{fume1n}), (\ref{fumeI}), obtained from the QH and the vanishing complexity factor condition are incompatible with the condition that the system tends asymptotically (as $t\rightarrow \infty$) to a static regime. On the other hand, in the case $B=1$ the QH condition implies that the shear scalar is function of $t$ only, and therefore in order to obtain the static asymptotic behavior, $\sigma$ should be function of both $t$ and $r$. 

\subsection{The asymptotic conditions}
In order to obtain the expected asymptotic behavior we  shall assume the same form of metric functions (\ref{fume1n}), (\ref{fumeI}), but replacing  $\sigma$ by an arbitrary  function of $t$ (say $f(t)$), such that  
  in the limit  $t\rightarrow \infty$
\begin{eqnarray}
F(t)\rightarrow\gamma=constant>0, \quad f(t)\rightarrow 0,\nonumber \\
 \frac{\dot f}{f^2}\rightarrow constant>0,
\label{asy1}
\end{eqnarray}
besides $g$ is not a constant.

Obviously such metric functions do not satisfy (\ref{Cond1}) (\ref{Cond2}) and  (\ref{exp4}),  in general (for any $t$) although they do satisfy such conditions in the limit $t\rightarrow \infty$.

From the comments above we shall assume  our metric variables to  read
\begin{eqnarray}
  A &=& \frac{\dot{f}}{2\beta^2 f^2} \left [2\beta^2-f^2 (\beta r+c_1)^2\right],\label{fume1}\\
  R &=& F(t) g(r)(\beta r+c_1)e^{\frac{f ^2 }{4\beta^2}(\beta r+c_1)^2}.\label{fume}
\end{eqnarray}

\noindent Using  these expressions  in (\ref{18}) and  (\ref{mu})-(\ref{fq}) we find  for the physical variables
{\footnotesize
\begin{eqnarray}\label{muvs}
  8\pi \mu=\frac{4\beta^4 f^4}{\dot f^2 [2\beta^2-f^2(\beta r+c_1)^2]^2}\left[\frac{\dot F}{F}+\frac{f \dot f(\beta r+c_1)^2}{2 \beta^2}\right]^2-2\left[\frac{g^{\prime \prime}}{g}+\frac{g^\prime \left[2 \beta^2+f^2(\beta r+c_1)^2\right]}{g \beta(\beta r+c_1)}\nonumber \right. \\ \left. +\frac{f^4(\beta r+c_1)^2}{4 \beta ^2}+\frac{3 f^2}{2}\right]-\left[ \frac{g^{\prime }}{g}+\frac{\beta}{(\beta r+ c_1)}+\frac{f^2 (\beta r+c_1)}{2\beta}\right]^2   +   \frac{e^\frac{{-f^2(\beta r+c_1)^2}}{2\beta^2}}{F^2 g^2 (\beta r+c_1)^2}
\end{eqnarray}
}
{\footnotesize
\begin{eqnarray}\label{Prvs}
  8\pi P_r=-\frac{4\beta^4 f^4}{\dot f^2\left[2\beta^2 -f^2(\beta r+c_1)^2\right]^2}\left\{\frac{2\ddot F}{F}+\frac{2 \dot F f \dot f (\beta r+c_1)^2}{F\beta^2}+\frac{(\beta r+c_1)^2(\dot f^2+f\ddot f)}{\beta^2}\right. \nonumber \\ \left.+\frac{\dot f^2 f^2(\beta r+c_1)^4}{2\beta^4}-2\left[\frac{\ddot f}{\dot f}-\frac{4\beta^2 \dot f}{f[2\beta^2-f^2(\beta r+c_1)^2]} \right] \left[\frac{\dot F}{F}+\frac{f\dot f (\beta r+c_1)^2}{2 \beta^2}\right]\right. \nonumber \\ \left.+\left[ \frac{\dot F}{F}+\frac{f\dot f (\beta r+c_1)^2}{2 \beta^2} \right]^2\right\} - \frac{e^\frac{{-f^2(\beta r+c_1)^2}}{2\beta^2}}{F^2 g^2 (\beta r+c_1)^2}-\frac{4\beta f^2(\beta r+c_1)}{[2 \beta^2-f^2(\beta r+c_1)^2]} \left[\frac{g^\prime}{g}+\frac{\beta}{(\beta r+c_1)}+\frac{f^2(\beta r+c_1)}{2\beta}\right]\nonumber  \\+\left[\frac{g^\prime}{g}+\frac{\beta}{(\beta r+c_1)}+\frac{f^2(\beta r+c_1)}{2\beta}\right]^2 \qquad \qquad
 \end{eqnarray}
}
{\footnotesize
\begin{eqnarray}\label{Ptvs}
  8\pi P_\bot=-\frac{4\beta^4 f^4}{\dot f^2\left[2\beta^2 -f^2(\beta r+c_1)^2\right]^2}\left\{\frac{\ddot F}{F}+\frac{ \dot F f \dot f (\beta r+c_1)^2}{F\beta^2}+\frac{(\beta r+c_1)^2(\dot f^2+f\ddot f)}{2\beta^2}\right. \nonumber \\ \left.+\frac{\dot f^2 f^2(\beta r+c_1)^4}{4\beta^4}  -\left[\frac{\ddot f}{\dot f}-\frac{4\beta^2 \dot f}{f[2\beta^2-f^2(\beta r+c_1)^2]} \right] \left[\frac{\dot F}{F}+\frac{f\dot f (\beta r+c_1)^2}{2 \beta^2}\right]\right\}\nonumber \\+\frac{g^{\prime \prime}}{g}-\frac{2\beta^2 f^2}{[2\beta^2-f^2(\beta r+c_1)^2 ]}+\frac{f^4(\beta r+c_1)^2}{4\beta^2} +\frac{g^\prime[2\beta^2+f^2(\beta r+c_1)^2]}{g\beta(\beta r+c_1)}-\nonumber \\ \frac{2\beta f^2(\beta r+c_1)}{[2 \beta^2-f^2(\beta r+c_1)^2]} \left[\frac{g^\prime}{g}+\frac{\beta}{(\beta r+c_1)}+\frac{f^2(\beta r+c_1)}{2\beta}\right]+\frac{3 f^2}{2}
\end{eqnarray}
}
{\footnotesize
\begin{eqnarray}\label{q}
  4\pi q=\frac{2\beta^2 f^2}{\dot f\left[2\beta^2 -f^2(\beta r+c_1)^2\right]}\left\{\frac{\dot F}{F}\left[\frac{g^\prime}{g}+\frac{\beta}{(\beta r+c_1)} +\frac{f^2(\beta r+c_1)}{2\beta} \right]+\frac{g^\prime f \dot f (\beta r+c_1)^2}{2\beta^2 g} \right. \nonumber \\ \left. +\frac{f \dot f(\beta r+c_1)}{2\beta}\left[3+\frac{f^2(\beta r+c_1)^2}{2\beta^2}\right]+\frac{2\beta f^2(\beta r+c_1)}{[2\beta^2-f^2(\beta r+c_1)^2]}\left[\frac{\dot F}{F}+\frac{f\dot f (\beta r+c_1)^2} {2\beta^2}  \right]   \right\}
  \end{eqnarray}
}
\begin{eqnarray}
m=\frac{R}{2}\left\{1+\frac{4R^2 \beta^4 f^4}{\dot f^2\left[2\beta^2 -f^2(\beta r+c_1)^2\right]^2}\left[ \frac{\dot F}{F}+\frac{f\dot f (\beta r+c_1)^2}{2 \beta^2} \right]^2 \nonumber \right. \\ \left.-R^2 \left[\frac{g^\prime}{g}+\frac{\beta}{(\beta r+c_1)}+\frac{f^2(\beta r+c_1)}{2\beta}\right]^2\right\}.
\label{mmod}
\end{eqnarray}

\subsection{The matching conditions}
So far our model is determined up to three functions $F(t), f(t), g(r)$. The form of these functions will be suggested by  the asymptotic conditions as $t \rightarrow \infty$ and  a condition to avoid shell crossing singularities ($R^\prime>0$).

We are looking for a model which asymptotically (as $t\rightarrow \infty$), approaches the state of a static ghost star $m(t\rightarrow \infty, r_{\Sigma^{(e)}})=0$.

As mentioned before,  for the required asymptotic behavior of the model,   we must demand  that  in the limit  $t\rightarrow \infty$, conditions (\ref{asy1}) are satisfied.

\noindent  Let us now consider the matching of this model on $\Sigma^{(e)}$. We shall demand the matching conditions (\ref{20}) and (\ref{20lum}) to be satisfied asymptotically  (as $t\rightarrow \infty$), when a ghost star is expected to form
Thus we shall demand
\begin{equation}
m( \infty, r_{\Sigma^{(e)}})=0.
\label{match1}
\end{equation}

On the other hand, as it can be seen from (\ref{q}) and (\ref{asy1}), in the limit $t\rightarrow \infty$, we obtain $q\rightarrow 0$ as expected from the static limit, therefore we also must demand
\begin{equation}
P_r (\infty, r_{\Sigma^{(e)}})=0.
\label{match1}
\end{equation}

Using (\ref{asy1}) in (\ref{mmod}), the  condition $m(t\rightarrow \infty, r_{\Sigma^{(e)}})=0$ reads
\begin{equation}
\gamma g_{\Sigma^{(e)}}(\beta r_{\Sigma^{(e)}}+c_1)\left[\frac{g^\prime_{\Sigma^{(e)}}}{g_{\Sigma^{(e)}}} +\frac{\beta}{(\beta  r_{\Sigma^{(e)}}+c_1)}\right]=1.
\label{match2}
\end{equation}

To specify further our model we shall assume for the function $g(r)$ and the constant $c_1$ 
\begin{equation}
g=c_3 r, \qquad c_1=0,
\label{match3}
\end{equation}
where $c_3$ is dimensionless constant.

Then condition (\ref{match2}) becomes
\begin{equation}
r_{\Sigma^{(e)}}=\frac{1}{2\gamma c_3 \beta}.
\label{match4}
\end{equation}

On the other hand condition (\ref{match1}) reads
\begin{equation}
\frac{g^\prime_{\Sigma^{(e)}}}{g_{\Sigma^{(e)}}} +\frac{1}{r_{\Sigma^{(e)}}}=\frac{1}{\gamma g_{\Sigma^{(e)}}\beta  r_{\Sigma^{(e)}}},
\label{match5}
\end{equation}
where (\ref{asy1}), and $c_1=0$ have been used.
Feeding back (\ref{match3}) into (\ref{match5}) we obtain 
\begin{equation}
r_{\Sigma^{(e)}}=\frac{1}{2\gamma c_3 \beta},
\label{match6}
\end{equation}
which is exactly  (\ref{match4}). Thus the above choice of constants ensure the asymptotic fulfillment of matching conditions on $\Sigma^{(e)}$, of our fluid distribution with Minkowski space-time.

It is worth mentioning that for this choice of $g$ and $c_1$ the matching condition are not satisfied on $\Sigma^{(i)}$. Therefore this model has a thin shell on this surface and $r_{\Sigma^{(i)}}$ is a free parameter.
\subsection{The model}
Finally, in order to fully describe our model we have to specify the two functions $F$ and $f$ which must satisfy  the  asymptotic conditions (\ref{asy1}). 

For the sake of simplicity we choose
\begin{equation}
F=\gamma e^{-\frac{r_{\Sigma^{(e)}}}{t}},\qquad f=-\frac{1}{t}.
\label{modf1}
\end{equation}

With the above choice, and (\ref{match3}) and (\ref{match6}), the metric functions  $A$ and $R$ read
\begin{equation}
A=1-\frac{x^2}{2t^{*2}},\qquad R=\frac{r_{\Sigma^{(e)}}}{2}e^{-1/t^*}x^2 e^{\frac{x^2}{4t^{*2}}},
\label{modf2}
\end{equation}
where $t^{*}\equiv\frac{t}{r_{\Sigma^{(e)}}}$ changing in the interval $[\frac{t_0}{r_{\Sigma^{(e)}}},\infty]$, with $t_0$ being a positive constant, and $x\equiv \frac{r}{r_{\Sigma^{(e)}}}$ changing in the interval $[\frac{r_{\Sigma^{(i)}}}{r_{\Sigma^{(e)}}},1]$. 

In order to ensure the positivity of $A$ we must assume $t^\ast>\frac{1}{\sqrt{2}}$. 

Using the above expressions in (\ref{muvs})--(\ref{mmod}), the physical variables describing our model read
\begin{equation}
8\pi \mu=\frac{4t^{\ast 4} (\frac{1}
{t^{\ast 2}}-\frac{x^2}{2 t^{\ast 3}})^2}{r^2_{\Sigma^{(e)}} (2t^{\ast 2}-x^2)^2}-\frac{2}{r^2_{\Sigma^{(e)}}}\left[\frac{(2t^{\ast 2}+x^2)}{t^{\ast 2}x^2}+\frac{x^2}{4t^{\ast 4}}+\frac{3}{2t^{\ast 2}}\right]-\frac{1}{r^2_{\Sigma^{(e)}}}\left(\frac{2}{x}+\frac{x}{2t^{\ast 2}}\right)^2+\frac{4 e^{(\frac{2}{t^\ast}-\frac{x^2}{2t^{\ast 2}})}}{r^2_{\Sigma^{(e)}} x^4},
\label{muf}
\end{equation}
\begin{eqnarray}
8\pi P_r=-\frac{4t^{\ast 4}} {r^2_{\Sigma^{(e)}}(2t^{\ast 2}-x^2)^2}\left\{ 2\left(\frac{1}{t^{\ast 4}}-\frac{2}{t^{\ast 3}}\right) -\frac{2x^2}{t^{\ast 5}}+\frac{3x^2}{t^{\ast 4}}+\frac{x^4}{2t^{\ast 6}} \nonumber \right. \\ \left. -2\left[-\frac{2}{t^\ast}+\frac{4t^\ast}{(2t^{\ast 2}-x^2)}\right]\left(\frac{1}{t^{\ast 2}}-\frac{x^2}{2t^{\ast 3}}\right)+\left( \frac{1}{t^{\ast 2}}-\frac{x^2}{2t^{\ast 3}}\right)^2\right\}\nonumber \\ -\frac{4 e^{(\frac{2}{t^\ast}-\frac{x^2}{2t^{\ast 2}})}}{r^2_{\Sigma^{(e)}} x^4}-\frac{4x}{r^2_{\Sigma^{(e)}}(2t^{\ast 2}-x^2)}\left(\frac{2}{x}+\frac{x}{2t^{\ast 2}}\right)+\frac{(\frac{2}{x}+\frac{x}{2t^{\ast 2}})^2}{r^2_{\Sigma^{(e)}}},
\label{prf}
\end{eqnarray}

\begin{eqnarray}
8\pi P_\bot=-\frac{4t^{\ast 4}} {r^2_{\Sigma^{(e)}}(2t^{\ast 2}-x^2)^2}\left\{ \left(\frac{1}{t^{\ast 4}}-\frac{2}{t^{\ast 3}}\right) -\frac{x^2}{t^{\ast 5}}+\frac{x^4}{4t^{\ast 6}}-\left[-\frac{2}{t^\ast}+\frac{4t^\ast}{(2t^{\ast 2}-x^2)}\right]\left(\frac{1}{t^{\ast 2}}-\frac{x^2}{2t^{\ast 3}}\right) \nonumber \right. \\ \left. +\frac{3 x^2}{2t^{\ast 4}}\right\}-\frac{2}{r^2_{\Sigma^{(e)}}(2t^{\ast 2}-x^2)}-\frac{2x}{r^2_{\Sigma^{(e)}}(2t^{\ast 2}-x^2)}\left(\frac{2}{x}+\frac{x}{2t^{\ast 2}}\right)+\frac{x^2}{r^2_{\Sigma^{(e)}}4t^{\ast 4}}\nonumber \\+\frac{(2t^{\ast 2}+x^2)}{r^2_{\Sigma^{(e)}}x^2t^{\ast 2}}+\frac{3}{2r^2_{\Sigma^{(e)}}t^{\ast 2}},
\label{ptf}
\end{eqnarray}
\begin{equation}
4\pi q=\frac{2}{r^2_{\Sigma^{(e)}}(2t^{\ast  2}-x^2)}\left[ \frac{2}{x}+\frac{x}{2t^{\ast 2}}-\frac{2x}{t^\ast}-\frac{x^3}{4t^{3\ast}}+\frac{x(2 t^\ast-x^2)}{t^\ast(2t^{\ast 2}-x^2)}\right],
\label{qf}
\end{equation}
\begin{equation}
m=\frac{r_{\Sigma^{(e)}}x^2 e^{-\frac{1}{t^\ast}}e^{\frac{x^2}{4t^{\ast 2}}}}{4}\left\{ 1+\frac{x^4 e^{-\frac{2}{t^\ast}} e^{\frac{x^2}{2t^{\ast 2}}}}{4} \left[ \frac{(2t{^\ast}-x^2)^2}{t^{\ast2} (2t^{\ast 2}-x^2)^2}-\frac{(4t^{\ast 2} +x^2)^2}{4x^2t^{\ast 4}}\right] \right\}.
\label{mf}
\end{equation}

The temperature for this model may be calculated using (\ref{V2}), producing
\begin{equation}
T^\ast =-\frac{1}{4\pi}\int{(\tau^\ast \frac{\partial q^\ast}{\partial t^\ast} +q^\ast A) dx}+\Phi(t),
\label{temp1}
\end{equation}
where $T^\ast \equiv \kappa Tr_{\Sigma^{(e)}}$, $\tau^\ast \equiv \frac{\tau}{r_{\Sigma^{(e)}}}$, $q^\ast \equiv 4\pi q r^2_{\Sigma^{(e)}}$, $\Phi(t)$ is an arbitrary function of integration, and $A$ and $q$ are given by (\ref{modf2}),  and (\ref{qf}), respectively.

However the resulting expression is cumbersome and not very illuminating. Still worse,  it depends on an arbitrary function ($\Phi$) and the numerical value of the relaxation time $\tau$. The former may be related to the temperature at the boundary surface, but this is also unknown unless we specify further the microphysics  of the fluid. On the other hand the numerical value of the relaxation time also depends on the microphysics  of the fluid. 
However a microscopic set up of the model is out of the scope of this work. 

Thus, the only way to  obtain the required information  is by assigning the value of $\tau$ and the profile of $\Phi$ in an {\it ad hoc} way, which  due to its intrinsic arbitrariness, deprives the obtained expression for the temperature  of any physical relevance. Accordingly we will dispense  with  temperature  graphic. 
Suffice is to say   that asymptotically the temperature tends to a constant  as expected from a static distribution in thermal equilibrium (as we shall see below the ``thermal inertial term''  $Ta$ \cite{it}    vanishes asymptotically  in this model).

We shall now illustrate the formation of the ghost star as  $t\rightarrow \infty$ for the model described so far. For doing that we need to evaluate the energy-density  in the limit  $t\rightarrow \infty$. Using (\ref{asy1}), (\ref{match3}) and (\ref{match6}) in (\ref{muvs}) we obtain

\begin{equation}
8\pi \mu(t\rightarrow \infty,r)=\frac{4}{r^2_{\Sigma^{(e)}}}\left(\frac{1-2x^2}{x^4}\right),
\label{gh1}
\end{equation}
where $x\equiv\frac{r}{r_{\Sigma^{(e)}}}$, whose values are within the interval $[\frac{r_{\Sigma^{(i)}}}{r_{\Sigma^{(e)}}},1]$.

\begin{figure}[h]
\includegraphics[width=4.in,height=6.in,angle=0]{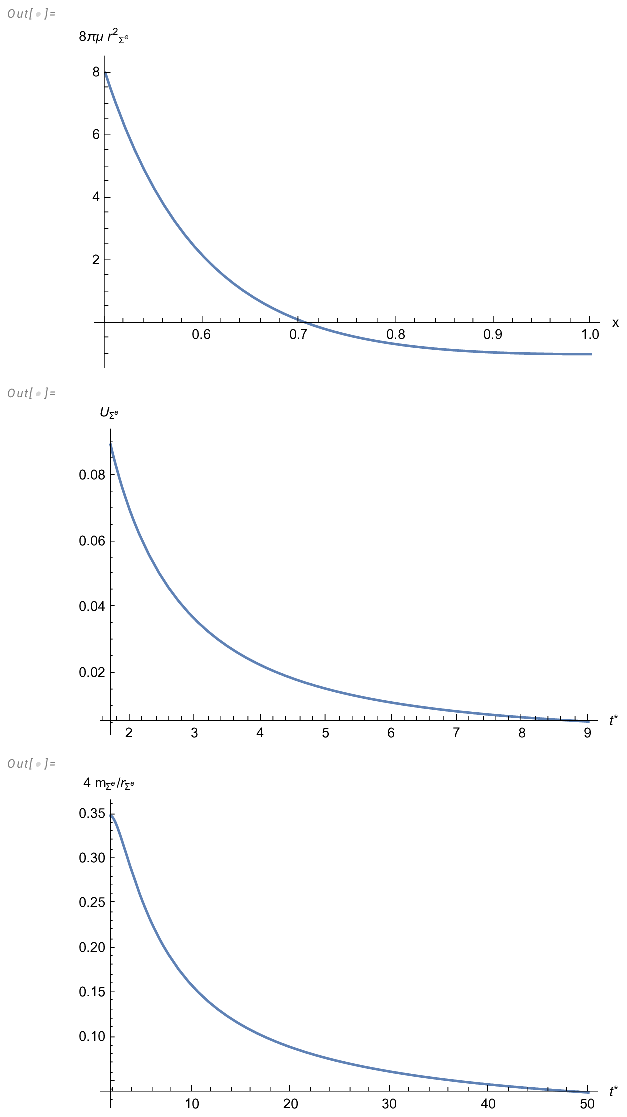}
\caption { $8\pi \mu  r^2_{\Sigma^{(e)}}$, evaluated at $t^\ast\rightarrow \infty$, as function of $x$ in the interval $[\frac{1}{2}, 1]$; $U_{\Sigma^{(e)}}$ and $m_{\Sigma^{(e)}}$ as functions of $t^\ast$.} \label{fig1b}
\end{figure}

\begin{figure}[h]
\includegraphics[width=4.in,height=6.in,angle=0]{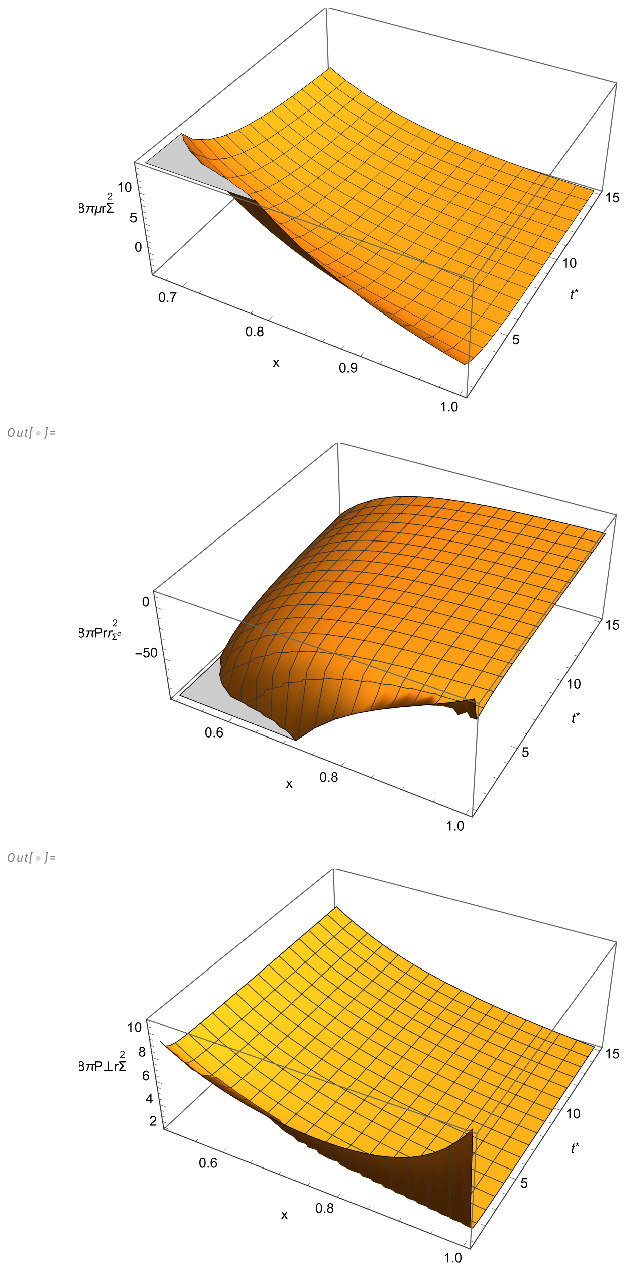}
\caption {\it  $8\pi \mu  r^2_{\Sigma^{(e)}}$, $8\pi P_r  r^2_{\Sigma^{(e)}}$ and $8\pi P_\bot  r^2_{\Sigma^{(e)}}$  as functions of $x$ and  $t^\ast$. \label{fig2b}} 
\end{figure}

\begin{figure}[h]
\includegraphics[width=4.in,height=6.in,angle=0]{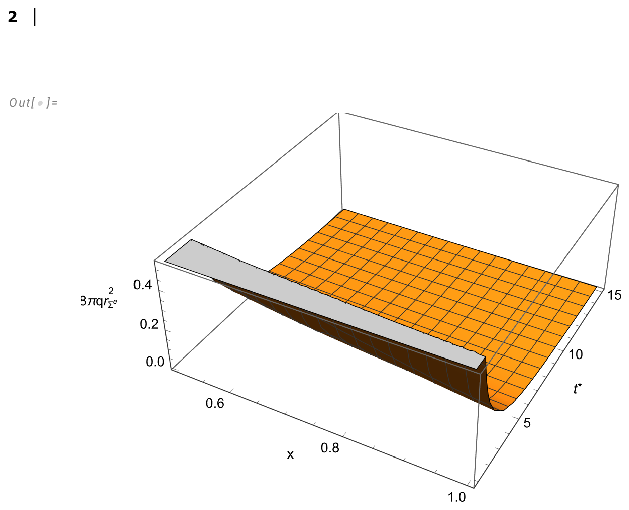}
\caption {\it  $8\pi q  r^2_{\Sigma^{(e)}}$  as function of $x$ and  $t^\ast$. \label{fig3b}} 
\end{figure}

With the choice above the expressions for $U_{\Sigma^{(e)}}$ and $m_{\Sigma^{(e)}}$ read 
\begin{equation}
U_{\Sigma^{(e)}}=\frac{(2t^\ast-1)\Psi}{2 t^\ast (2t^{2\ast}-1)},
\label{modf3}
\end{equation}
and
\begin{equation}
m_{\Sigma^{(e)}}=\frac{\Psi r_{\Sigma^{(e)}}}{4}\left[1+\frac{\Psi^2 (1-\frac{1}{2t^\ast})^2}{t^{\ast 4} (2-\frac{1}{t^{2\ast}})^2} -\frac{\Psi^2}{4}(2+\frac{1}{2 t^{\ast 2}})^2\right],
\label{modf3}
\end{equation}
where $\Psi\equiv e^{-\frac{1}{t^\ast}+\frac{1}{4t^{\ast 2}}}$.

The three curves in  Figure ~\ref{fig1b} illustrate the emergence of the ghost star. The first curve depicts the  radial distribution of energy-density as $t\rightarrow \infty$, which shows a  region of negative values of this variable, which ensures the vanishing of the total mass, as illustrated by the third curve. Finally the second curve shows the tendency to the static situation.

The behavior of the physical variables is depicted in Figure ~\ref{fig2b} and Figure ~\ref{fig3b}. The graphics for $P_r, P_\bot$ and $q$ have been drawn for the variables $t^\ast$ and $x$ in the intervals $[.9, 15]$ and $[.5, 1]$ respectively. Instead, for $\mu$ we have chosen the intervals $[1.7, 15]$ and $[.67, 1]$ in order to better illustrate the appearance of the region of negative energy-density. The fast convergence of the system to the static regime is well illustrated in Figure ~\ref{fig3b}.

\section{Discussion}

The main purpose of this work has been to exhibit the viability of the formation  of a ghost star as the end point of the evolution of a self--gravitating fluid distribution. To  achieve this goal we have presented an analytical  model of a dissipative spherically symmetric fluid distribution evolving toward a ghost star.  

We have initially found a primeval solution satisfying the vanishing complexity factor condition,  the quasi--homologous evolution and $B=1$. This primeval solution was next modified to satisfy the required asymptotic conditions (\ref{asy1}). Finally we have chosen the remaining arbitrary functions to fully specify our model. This final model represents and expanding fluid distribution  endowed with a cavity surrounding the center, tending to a static configuration. The endpoint of the evolution of this model is a ghost star as illustrated by Figure 1.

Furthermore, in the limit $t\rightarrow \infty$ we  have $A\rightarrow 1$, implying, because of  (\ref{5c}), the vanishing of the four-acceleration of the fluid forming the ghost star (which explains the vanishing of the ``thermal inertial term''  mentioned before). This implies in its turn, according to (\ref{28}), that the gravitational term in the dynamic equation (\ref{3m}) (the Tolman mass \cite{agm}) vanishes, and the equilibrium is reached by the balance between the radial pressure gradient and the anisotropic factor. A particular model of a ghost star with vanishing complexity factor and vanishing Tolman mass has been  considered in \cite{1n}.

The model satisfies asymptotically  Darmois conditions on the external boundary surface, whereas on the inner boundary surface such conditions are not satisfied  indicating thereby the appearance of shells on this  hypersurface.  The presence of  these  thin  shells are likely to be produced by the simplicity of the model.  More involved analytical models, or  numerical models could avoid these ``drawbacks''.

We should recall that the very existence of ghost stars relies on the assumption of the existence of regions of the fluid distribution endowed with negative energy-density.  In this respect it should be mentioned that negative energy-density (or negative mass) is a subject extensively considered in the literature (see \cite{bon}-\cite{bor} and references therein). Particularly relevant are those references relating the appearance of negative energy-density with quantum effects.

It is worth mentioning the relevance of the observational aspects of the ghost stars, in general and of our model in particular. On the one hand it is evident that the shadow of such kind of object should differ from the one produced by a self-gravitating star with non-vanishing total mass. In the particular case of the model here considered one should be able to detect the variation of the shadow as the system approaches the state of ghost star. We ignore if the ongoing observations of this kind  \cite{et1}-\cite{et4} are able to do that, but this is an important issue to consider. A research endeavor   pointing in a similar direction has been recently published in  \cite{nc}.

In the same order of ideas, it should be clear that the radiation emitted from the surface of a a ghost star should  not  exhibit  gravitational redshift, opening  the way for a possible detection of such objects. In our case a continuous measuring of such redshift, and the ensuing decreasing along the evolution, would indicate the formation of a ghost star.

On the other hand, it is worth noticing that ghost stars are a sort of reservoirs of dark mass produced by  the appearance of negative energy-density is some regions of the fluid distribution. It remain to be seen if the general problem of dark matter could be, at least partially, be explained in terms of ghost stars \cite{rep}.

We would also like to mention  that an  important piece of theoretical evidence behind the concept of ghost star is still missing. We have in mind a microscopic theory accounting for the appearance of negative energy-density. Research in this direction could provide further support to the astrophysical relevance of  ghost stars, by allowing to clarify important questions about the structure of these objects,  such as its stability.

Finally, let us mention two natural extensions of the work here presented
\begin{itemize}
\item Our solution was based on a set of heuristic conditions mentioned above. Alternatively,  solutions of this kind might  be found by using the general methods presented in \cite{TM, TMb, ivanov1, ivanov}, or resorting to some of  recently presented  results in the study of gravitational collapse (see for example \cite{nr1,nr2,nr3} and references therein).
\item We have resorted to GR to describe the gravitational interaction. It would be interesting  to consider the same problem within the context of one of the extended gravitational theories \cite{ext}.
\end{itemize}
\section{Appendix}

Dynamical equation.

Using (\ref{5c}), (\ref{18}),  (\ref{19}) and  (\ref{Prvs}) it can be easily obtained 
\begin{equation}
D_TU=-\frac{m}{R^2}-4\pi  P_r R,
+Ea, \label{28}
\end{equation}

which allows to write the dynamical equation following  from the Bianchi identities as (see \cite{7n} for details)
\begin{eqnarray}
\left( \mu+ P_r\right)D_TU
=-\left(\mu+ P_r \right)
\left[\frac{m}{R^2}
+4\pi  P_r R\right]
\nonumber \\-E^2\left[D_R  P_r
+2(P_r-P_{\perp})\frac{1}{R}\right]
-E\left[D_T q+2 q\left(2\frac{U}{R}+\sigma\right)\right].
\label{3m}
\end{eqnarray}

\vspace{6pt}

\section{Acknowledgemets}{This work was partially supported by the Grant PID2021-122938NB-I00 funded
by MCIN/AEI/ 10.13039/501100011033 and by ERDF A way of making Europe,
as well as the Consejer\'\i a de Educaci\'on of the Junta de Castilla y Le\'on under the
Research Project Grupo de Excelencia  Ref.:SA097P24 (Fondos Feder y en
l\'\i nea con objetivos RIS3).}

\end{document}